\documentclass[epsfig,onecolumn,floats,showpacs]{revtex4}
\usepackage{graphicx}
\usepackage{dcolumn}
\usepackage{bm}
\usepackage{amssymb}
\usepackage{amsmath}
\usepackage{epsfig}
\usepackage{subfigure}

\begin{document}

\title{Quantum simulation of Heisenberg spin chains with next nearest neighbor interactions in coupled cavities}
\author{Zhi-Xin Chen}
\author{Zheng-Wei Zhou}
\email{zwzhou@ustc.edu.cn}
\author{Xingxiang Zhou}
\email{xizhou@ustc.edu.cn}
\author{Xiang-Fa Zhou}
\author{Guang-Can Guo}
\affiliation{Key Laboratory of Quantum Information, University of
Science and Technology of China, Chinese Academy of Sciences,
Hefei, Anhui 230026, China}

\begin{abstract}
  We propose a scheme to simulate one-dimensional XXZ-type Heisenberg
  spin models with competing interactions between nearest-neighbors
  (NNs) and next-NNs in photon-coupled micro-cavities. Our scheme, for
  the first time, exploits the rich resources and flexible controls
  available in such a system to realize arbitrarily adjustable ratios
  between the effective NN and next-NN coupling strengths. Such
  powerful capability allows us to simulate frustration phenomena and
  disorder behaviors in 1-d systems arising from next-NN interactions,
  a large class of problems of great importance in condensed matter
  physics. Our scheme is robust due to the lack of atomic excitations which
  suppresses spontaneous emission and cavity decay strongly.

\end{abstract}


\maketitle

Quantum simulation is an important application of quantum
information science. As a promising physical system for quantum
simulation, ultracold atoms trapped in an optical lattice
\cite{Jaksch, Greiner} have many appealing properties such as long
coherence times, the possibility of simultaneous initialization of a
large number of particles, and tunable effective coupling strength
within a large range. However, it is difficult to focus a laser beam
on a single atom due to the short optical lattice period which makes
it challenging to realize single-site operations \cite {Calarco}. To
address this challenge, recently arrays of photon-coupled
microcavities have been suggested as an alternative quantum
simulator \cite{Plenio1,Plenio, Greentree,
  Angelakis, Hartmann, Fazio, Liu, Bose1, Sun, Kay, Bose2, Cho, Zhou, Splillance, Hennessy, Wallraff}. An
artificial system on a micro-chip, microcavities can be fabricated
in desired structures and dimensions with great precision, a
significant advantage that allows to realize both single-site
operations and neighboring-sites interactions easily \cite{Plenio1}.

In previous research on quantum simulation, the focus has been on
realizing various spin models with on-site and NN interactions,
the most widely used model in studies of condensed-matter physics
problems such as quantum magnetism \cite{Schollwock,Sachdev}. For
  instance, it was demonstrated that an array of coupled cavities with one atom in each cavity
  can be used to simulate the anisotropic (XYZ) Heisenberg spin-$\frac1 2$
  model. It was further shown that the XXZ-type Heisenberg spin model of any high
  spin can be realized using a cavity array with a number of atoms
  in each cavity \cite{Cho}. The XXZ-type Heisenberg spin
  model can also be realized when the spin states are represented by polaritons \cite{Liu, Kay}.
  Much of the recent research was reviewed in \cite{Plenio1}. However, it is well known
in condensed-matter physics that many important physics and exotic
phenomena arise due to the long-range nature of interactions between
spins. A particularly important model to capture this is a spin model
with both (and possibly competing) NN and next-NN interactions, which
is known to reveal many important physics such as Tomonaga-Luttinger
liquid states and spin-Peierls states \cite{Sachdev, Haldane,
  Somma}. It is even suggested that it could be used to study the
mechanism for iron-based superconductors \cite{Zhao}. The
  importance of such frustrated spin systems has also been recognized
  in the context of quantum information. For instance, behavior of
  quantum entanglement in frustrated spin systems was studied by a
  number of authors \cite{Vedral, Nielsen, Lewenstein, Wang, Sun2} and
  interesting and deep connections between entanglement and phase
  transitions were dicovered in these systems. Unfortunately, physical
  implementation for a spin-chain quantum
simulator with both tunable NN and next-NN interactions has not
been available, mainly due to the technical difficulty in
realizing controllable next-NN interactions. For instance, in
ultracold atomic systems the adjustable spin interactions are
usually realized by controlling the wave function overlap between
neighboring sites. This technique is not useful in engineering
next-NN interactions since the wavefunction overlap falls off
exponentially with distance.

In this work, we show how one can implement an effective spin model in
a microcavity system with both NN and next-NN interactions. By taking
advantage of the many controls available and using a smart idea of
interaction cancellation and enhancement, we can adjust at will the
ratio between the NN and next-NN interaction strengths.  This then
allows for the first time to simulate a large class of
condensed-matter physics problems in which the competition between NN
and next-NN interactions plays an essential role. Moreover, in our
approach, cavity field is always kept in ground state and atoms remain
in two long-lived states. Since excited atomic levels and cavity
photon modes only appear in virtual processes, spontaneous emission of
internal states and cavity decay are strongly suppressed.

\begin{figure}[tbp]
\centerline{\includegraphics[height=.4\textheight]{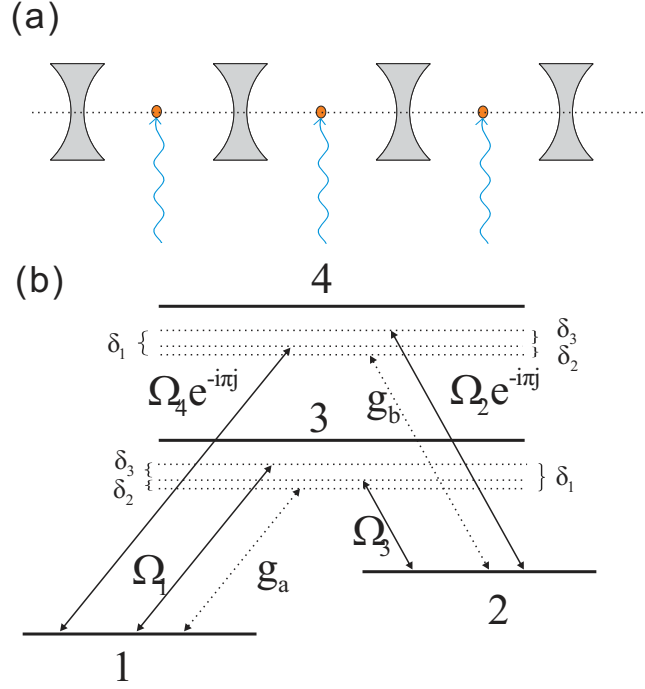}}
\caption{(color online). (a) A one-dimensional array of coupled
cavities with one 4-level atom in each cavity. (b) Involved atomic
levels and transitions.} \label{fig:simple}
\end{figure}

We consider an array of cavities that are coupled via exchange of
photons with one 4-level atom in each cavity (Fig.1). Such a model can
be realized in several kinds of physical systems such as microtoroidal
cavity arrays \cite{Splillance}, photonic crystal defects
\cite{Hennessy} and superconducting stripline resonators
\cite{Wallraff}. Two long-lived levels, $|1\rangle $ and $|2\rangle$,
represent the two spin states $|\downarrow \rangle $ and $%
|\uparrow \rangle $ for the effective spin. Together with two
excited states $|3\rangle$ and $|4\rangle$, they form two
independent $\Lambda$ level structures. We denote the subsystem
consisting of levels $|1\rangle$, $|2\rangle$, and $|3\rangle$
$\Lambda_a$, and that consisting of $|1\rangle$, $|2\rangle$, and
$|4\rangle$ $\Lambda_b$.

In previous research, a simpler atomic level configuration with one
$\Lambda$ structure was used to realize effect spin models with NN XXZ
interactions \cite{Hartmann,Cho}. In these models, interaction
strength between spins decreases rapidly with distances and next-NN
interactions are negligibly small compared to NN interactions. In our
study, by cleverly adjusting relevant experimental conditions, we can
make interactions arising due to the two independent level structures
$\Lambda_a$ and $\Lambda_b$ add up or cancel each other depending on
the phases of the control lasers. This is the key idea that allows us
to realize arbitrary ratios between NN and next-NN interaction
strengths.

We now derive the effective Hamiltonian of the photon coupled
microcavity system in Fig. 1 (a) when each atom couples with two
cavity modes and four external lasers (Fig.1(b)). Suppose two cavity
modes, $\omega_a$ and $\omega_b$, are close in energy to the
transitions $|1\rangle \leftrightarrow |3\rangle$ and $|2\rangle
\leftrightarrow |4\rangle$, and thus drive these two transitions with
strengths $g_a$ and $g_b$ respectively. Further, we apply four lasers
with frequencies $\omega_1$, $\omega_2$, $\omega_3$ and $\omega_4$ to
drive the transitions $|1\rangle\leftrightarrow |3\rangle$, $|2\rangle
\leftrightarrow |4\rangle$, $%
|2\rangle \leftrightarrow |3\rangle$ and $|1\rangle \leftrightarrow
|4\rangle $, with Rabi frequencies $\Omega_1$, $\Omega_2 e^{-i\pi
j}$, $\Omega_3$, $\Omega_4 e^{-i\pi j}$ (the index $j$ represents
the $j$th cavity), respectively(without loss of generality, here,
assume all the $\Omega_i$ are real). Here, we have modulated the
phases of lasers $\omega_2$ and $\omega_4$ on purpose, and as will
be seen later this phase modulation plays a key role in our scheme.
All the transitions are assumed to be large detuned, i.e., the
magnitudes of the detunings $\delta_{31} =\omega_{31} -\omega_a$,
$\delta_{42} =\omega_{42} -\omega_b$, $\Delta_{31} =\omega_{31}
-\omega_1$, $\Delta_{42} =\omega_{42} -\omega_2$, $\Delta_{32}
=\omega_{32} -\omega_3$, and $\Delta_{41} =\omega_{41} -\omega_4$ (
$\omega_{\mu\nu}$ is the energy difference between level
$|\mu\rangle$ and $|\nu\rangle$) are much greater than the
transition strengths $|g_a|$, $|g_b|$ and $|\Omega_i|$$
(i=1,2,3,4)$. Under these conditions, in the rotating frame, we can
perform a standard adiabatic elimination of the atomic excited
states $|3\rangle$ and $|4\rangle$ \cite{James} and obtain an
effective Hamiltonian

\begin{equation}
\begin{split}
H=&-\sum_{j}[A_1|1_j\rangle\langle 1_j| a_j
e^{i\delta_1t}+A_2|2_j\rangle\langle 1_j| a_j e^{i\delta_2 t}+H.c.] \\
&-\sum_{j}[(-1)^j B_1|2_j\rangle\langle 2_j| b_j e^{i\delta_1 t}+(-1)^j
B_2|1_j\rangle\langle 2_j| b_j e^{i\delta_2 t}+H.c.] \\
&-\sum_{j}[(A_3|1_j\rangle\langle 2_j| +B_3|2_j\rangle\langle
1_j|)e^{i\delta_3 t}+H.c. ] \\
&-\sum_{j}[(\frac{\Omega_1^2}{\Delta_{31}}+\frac{\Omega_4^2}{\Delta_{41}})|1_j\rangle\langle
1_j|+
(\frac{\Omega_3^2}{\Delta_{32}}+\frac{\Omega_2^2}{\Delta_{42}})|2_j\rangle\langle 2_j|] \\
&-\sum_{j}(\frac{g_a^2}{\delta_{31}}|1_j\rangle\langle 1_j|a_j^\dagger a_j+%
\frac{g_b^2}{\delta_{42}}|2_j\rangle\langle 2_j|b_j^\dagger b_j) \\
&+\sum_{j}[J_a(a_j^\dagger a_{j+1}+a_j a_{j+1}^\dagger)+J_b(b_j^\dagger
b_{j+1}+b_j b_{j+1}^\dagger)],
\label{eq:first_Heff}
\end{split}
\end{equation}
where $\delta_1=\delta_{31}-\Delta_{31}=\delta_{42}-\Delta_{42}$, $%
\delta_2=\delta_{31}-\Delta_{32}=\delta_{42}-\Delta_{41}$, and $%
\delta_3=\delta_1-\delta_2$ are the two-photon detunings,
$A_1=\frac{
  \Omega_1 g_a}{2}(\frac{1}{\Delta_{31}}+\frac{1}{\delta_{31}})$,
$A_2=\frac{ \Omega_3
  g_a}{2}(\frac{1}{\Delta_{32}}+\frac{1}{\delta_{31}})$, $A_3=\frac{
  \Omega_1 \Omega_3}{2}(\frac{1}{\Delta_{31}}+\frac{1}{\Delta_{32}})$,
$B_1= \frac{\Omega_2
  g_b}{2}(\frac{1}{\Delta_{42}}+\frac{1}{\delta_{42}})$, $B_2=
\frac{\Omega_4 g_b}{2}(\frac{1}{\Delta_{41}}+\frac{1}{\delta_{42}})$,
and $B_3= \frac{\Omega_2
  \Omega_4}{2}(\frac{1}{\Delta_{41}}+\frac{1}{\Delta_{42}})$ are the
effective coupling strengths, $J_a$, $J_b$ are the tunnelling rate
of photons between neighboring cavities(all assumed to be real).
Notice that the effective interaction strengths $A_1$ and $A_2$
are induced by the level structure $\Lambda_a$, whereas $B_1$ and
$B_2$ arise due to $\Lambda_b$, and there is a phase factor of
$(-1)^j$ from the Rabi frequencies $\Omega_2 e^{-i\pi j}$ and
$\Omega_4 e^{-i\pi j}$ in the $j$th cavity.

We can now derive the effective spin interactions using the
virtual-transition induced by effective Hamiltonian in Eq.
(\ref{eq:first_Heff}). To avoid excitations of real photons and
ensure that all two-photon transitions are independent when we
derive the effective spin Hamiltonian \cite{Cho}, we further
require that all two-photon transitions be large detuned, i.e.,
$|\delta _i|$, $|\delta _3-\delta _2|$ $\gg $ $|A_i|$, $|B_i|$,
$|J_a|$, $|J_b|$, $|\frac{g_a^2}{\delta
  _{31}}|$, $|\frac{g_b^2}{\delta _{42}} |$(i=1,2,3). We define the
spin operators $S_j^z=\frac 12(|2_j\rangle \langle
2_j|-|1_j\rangle \langle 1_j|)$, $S_j^{+}=|2_j\rangle \langle
1_j|$, $S_j^{-}=|1_j\rangle \langle 2_j|$. Before proceeding, we
simplify the Hamiltonian in Eq. (\ref{eq:first_Heff}). First, when
we consider the third term up to second order, it is $\sum_j \frac
{2}{\delta_3} (B_3^2-A_3^2) S_j^z$, so the third term and the
fourth term only contribute to effective local magnetic field. We
temporarily neglect them.  Second, the fifth term is a
perturbation term that modifies the detunings of the two-photon
transitions, i.e., $\delta _{a1}=\delta _1+\frac{g_a^2}{\delta
_{31}}$, $\delta _{a2}=\delta _2+\frac{g_a^2}{\delta _{31}}$,
$\delta _{b1}=\delta
_1+\frac{g_b^2}{\delta _{42}}$, $\delta _{b2}=\delta _2+\frac{%
  g_b^2}{\delta _{42}}$. Finally, we assume periodic boundary
conditions and take advantage of Fourier transformation to
diagonalize the photon coupling terms. Defining $a_j(b_j)=\frac
1{\sqrt{N}}\sum_{k=1}^N F_{jk}c_k(d_k)$, where $N$ is the total
number of the cavities, $F_{jk}=exp(-i\frac{2\pi } Njk)$ and
$\sum_{j=1}^N F_{jk}^* F_{jl}=N \tilde{\delta}_{kl}$
($\tilde{\delta}_{kl}$ is the Kronecker function), we have

\begin{equation}
\sum_jJ_a(a_j^{\dagger }a_{j+1}+a_ja_{j+1}^{\dagger
})=\sum_kT_{ak}c_k^{\dagger }c_k,
\end{equation}
\begin{equation}
 \sum_jJ_b(b_j^{\dagger
}b_{j+1}+b_jb_{j+1}^{\dagger })=\sum_kT_{bk}d_k^{\dagger }d_k,
\end{equation}
where $T_{ak}=J_a \sum_{jl} \frac 1N (F_{jk}^*F_{j+1,l}+F_{jk}
F_{j+1,l}^*)=2J_a cos(\frac{2\pi}{N}k)$ and $T_{bk}=2J_b
cos(\frac{2\pi}{N}k)$. The Hamiltonian in the rotating frame reads
\begin{equation}
\begin{split}
H=& -\sum_{j,k}\{[A_1(\frac 12I_j-S_j^z)e^{i\delta
_{a1}t}+A_2S_j^{+}e^{i\delta _{a2}t}]\frac{F_{jk}}{\sqrt{N}}c_k e^{-iT_{ak}}+H.c.\}  \\
& -\sum_{j,k}\{(-1)^j[B_1(\frac 12I_j+S_j^z)e^{i\delta
_{b1}t}+B_2S_j^{-}e^{i\delta
_{b2}t}]\frac{F_{jk}}{\sqrt{N}}d_k e^{-iT_{bk}}+H.c.\}.  \\
\end{split}
\end{equation}
When the large detuning conditions $|\delta
_{ai}|$, $|\delta _{bi}|$, $|\delta _{a1}-\delta _{a2}|$%
, $|\delta _{b1}-\delta _{b2}|$ $\gg $ $|A_i|$, $|B_i|$, $|J_a|$,
$|J_b|$ are satisfied, all terms are independent when we
adiabatically eliminate the photon states, and we obtain the
effective spin Hamiltonian

\begin{equation}
\begin{split}
H=& \sum_{j,l,k}[\frac 1N \frac{F_{jk}F_{lk}^*}{\delta_{a1}-T_{ak}}
A_1^2 (\frac
12I_j-S_j^z)(\frac 12I_l-S_l^z)+\frac 1N \frac{F_{jk}F_{lk}^*}{{\delta_{a2}-T_{ak}}}A_2^2 S_j^{+}S_l^{-}] \\
&+\sum_{j,l,k}(-1)^{j+l}[\frac 1N
\frac{F_{jk}F_{lk}^*}{\delta_{b1}-T_{bk}}B_1^2(\frac
12I_j+S_j^z)(\frac 12I_l+S_l^z)+\frac 1N \frac{F_{jk}F_{lk}^*}{{\delta_{b2}-T_{bk}}}B_2^2S_j^{-}S_l^{+}].  \\
\end{split}
\end{equation}
It is clear from Eq. (5) that effective spin interactions arise
between any pair of lattice sites $j$ and $l$ due to nonlocal
modes $c_k$ and $d_k$. However, due to the large detuning
conditions $|\delta_{\mu i}|\gg|T_{\mu k}|=|2J_{\mu} cos(\frac
{2\pi}N k)|$ $(\mu=a,b;i=1,2)$, the interaction strength drops
quickly with the site distance $|i-j|$. To see this, we expand the
following term for interaction strengths in Eq. (5) to second
order in $T_{\mu k} / \delta_{\mu i}$. Making use of the relations
$\sum_k F_{jk} F_{lk}^*=N \tilde{\delta}_{jl}$, $\sum_k F_{jk}
F_{lk}^*\cdot 2cos(\frac {2\pi} N k )=N
(\tilde{\delta}_{j,l+1}+\tilde{\delta}_{j+1,l})$, $\sum_k F_{jk}
F_{lk}^*\cdot (2cos(\frac {2\pi} N k ))^2=N(2\tilde {\delta}_{jl}+
\tilde{\delta}_{j,l+2}+\tilde{\delta}_{j+2,l})$, we obtain
\begin{equation}
\begin{split}
\frac1N\sum_k \frac{F_{jk}F_{lk}^*}{{\delta_{\mu i}-T_{\mu
k}}}&\approx \frac 1N \sum_k\frac{F_{jk}F_{lk}^*}{\delta_{\mu
i}}(1+\frac{T_{\mu k}}{\delta_{\mu
i}}+\frac{T_{\mu k}^2}{\delta_{\mu i}^2})\\
&= \frac{1}{\delta_{\mu i}}(1+\frac{2J_\mu^2}{\delta_{\mu i}^2})
\tilde{\delta}_{j,l}+\frac{J_\mu}{\delta_{\mu
i}^2}(\tilde{\delta}_{j,l+1}+\tilde{\delta}_{j+1,l})+\frac{J_\mu^2}{\delta_{\mu
i}^3}(\tilde{\delta}_{j,l+2}+\tilde{\delta}_{j+2,l}).
\end{split}
\end{equation}
Substituting the expanded terms in Eq. (5), we obtain the following
effective spin Hamiltonian to second order in $J_{\mu} /\delta_{\mu
i}$:
\begin{equation}
\begin{split}
H=&\sum_j \{2[\frac{J_a}{\delta_{a1}^2} A_1^2+(-1)^1
\frac{J_b}{\delta_{b1}^2}B_1^2] S_j^z S_{j+1}^z \\
&+[\frac{J_a}{\delta_{a2}^2} A_2^2+(-1)^1 \frac{J_b}{\delta_{b2}^2}
B_2^2](S_j^+ S_{j+1}^- + S_j^- S_{j+1}^+)\\
&+2[\frac{J_a^2}{\delta_{a1}^3} A_1^2+(-1)^2
\frac{J_b^2}{\delta_{b1}^3}B_1^2] S_j^z S_{j+2}^z \\
&+[\frac{J_a^2}{\delta_{a2}^3} A_2^2+(-1)^2
\frac{J_b^2}{\delta_{b2}^3} B_2^2](S_j^+ S_{j+2}^- +S_j^- S_{j+2}^+)\\
&+h_j S_j^z\},
\end{split}
\end{equation}
where $h_j$ is the strength of local effective magnetic field
\cite{mag}.  It can be see from Eq. (7) that the effective spin
interaction strength terms proportional to $A_i$ and $B_i$ arise from
the atom's level structures $\Lambda_a$ and $\Lambda_b$
separately. For both $\Lambda_a$ and $\Lambda_b$, the next-NN
interaction strength is 1 order of magnitude (in $J/\delta$) smaller
than that of NN interaction. Therefore, if only one of the level
structures $\Lambda_a$ and $\Lambda_b$ was present, it can be seen
from the above Hamiltonian that the next-NN interaction would have
been negligibly weak compared to NN interaction because their ratio is
$J_\mu / \delta_{\mu i} \ll 1$. For this reason, the next-NN
interactions are omitted in previous work \cite{Hartmann,Cho}.
However, in our system both level structures $\Lambda_a$ and
$\Lambda_b$ contribute to the effective spin interactions. When
calculating the total effective interaction strength, we see that
contributions due to $\Lambda_a$ and $\Lambda_b$ tend to cancel for NN
interactions but add up for next-NN interactions, due to our careful
choice of the control laser phases. By using this smart idea of
interaction cancellation and enhancement, we can arbitrarily adjust
the ratio between NN and next-NN interaction strengths. Finally, we
simplify the above effective spin Hamiltonian and write it in the
following form

\begin{equation}
\begin{split}
H=\sum_j \left[ \left[ \sum_{\sigma =1,2}J_\sigma
(S_j^xS_{j+\sigma }^x+S_j^yS_{j+\sigma }^y)+\lambda _\sigma
S_j^zS_{j+\sigma }^z \right]+ h_jS_j^z \right]. \label{eq:final_H}
\end{split}
\end{equation}
Here, the effective coupling strengths are $J_1=2(\frac{J_a}{\delta
  _{a2}^2} A_2^2-\frac{J_b}{\delta _{b2}^2}B_2^2)$,
$J_2=2(\frac{J_a^2}{\delta _{a2}^3} A_2^2+\frac{J_b^2}{\delta
  _{b2}^3}B_2^2)$, $\lambda _1=2(\frac{J_a}{\delta
  _{a1}^2}A_1^2-\frac{J_b}{\delta _{b1}^2}B_1^2)$, $\lambda
_2=2(\frac{J_a^2}{ \delta _{a1}^3}A_1^2+\frac{J_b^2}{\delta
  _{b1}^3}B_1^2)$.

The interaction cancellation and enhancement is clearly reflected in
the above expressions for $J_1$ and $J_2$, effective interaction
strengths between NN and next-NN spins in the Hamiltonian
Eq. (\ref{eq:final_H}). Obviously, by choosing close values for
$\frac{J_a}{\delta_{a_i}^2} A_i^2$ and $\frac{J_b}{\delta
  _{b_2}^2}B_i^2$$(i=1,2)$, we can make $J_2$ comparable to or even
greater than $J_1$. We can also adjust the relative signs of $J_1$ and
$J_2$ easily.

\begin{figure}[tbp]
\centerline{\includegraphics[height=.4\textheight]{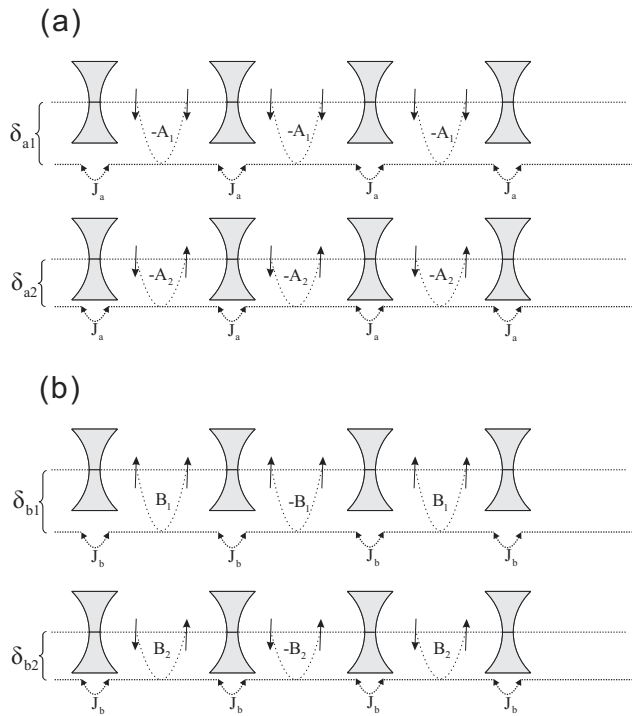}}
\caption{Atoms in cavities interact via exchange of virtual photons of
  cavity modes $\omega_a$ ((a)) and $\omega_b$ ((b)). The two spin
  states $|\downarrow \rangle $ and $%
  |\uparrow \rangle $ are represented by two long-lived atomic levels
  $|1\rangle $ and $|2\rangle$. $A_i$, $B_i(i=1,2)$ are the effective
  coupling strengths, $\delta_{ai}$, $\delta_{bi}$ are the detunings,
  $J_a$, $J_b$ are the tunnelling rates of photons between neighboring
  cavities.  } \label{fig:simple}
\end{figure}

To gain a better intuitive understanding of the critical role that
interaction cancellation and enhancement plays in the physics
underlying the effective Hamiltonian in Eq. (\ref{eq:final_H}), we
present a simple physical picture in Fig.2. Fig. 2 (a) describes
interactions induced by the two-photon transitions of the $\Lambda_a$
level structure. In this process, atoms interact via the exchange of
virtual photons of the cavity mode $\omega_a$.  Similarly, Fig.2(b)
describes interactions induced by the two-photon transitions of the
$\Lambda_b$ level structure, where atoms interact via the exchange of
virtual photons of the cavity mode $\omega_b$.  The difference is, due
to the phase modulation of the control lasers, in Fig.2(b), the
amplitude of emitting or absorbing a photon is site-dependent, i.e.,
$(-1)^{j+1}B_i$ $(i=1,2)$ in the $j$th cavity.

Now, we consider the first case in Fig.2(a), which generates ZZ
interactions. The probability amplitude that the first atom emits a
virtual photon with frequency $\omega_a$ is $%
\frac{-A_1}{\delta_{a1}}$, where $\delta_{a1}$ is the detuning. Then
the photon tunnels to the second cavity with amplitude
$\frac{J_a}{\delta_{a1}}$. If it is absorbed with amplitude $(-A_1)$,
the strength of NN interaction is $\frac{-A_1}{\delta_{a1}%
} \frac{J_a}{\delta_{a1}} (-A_1)=\frac{J_a}{\delta _{a1}^2}A_1^2>0$.
The photon can further tunnel to the third cavity with probability
amplitude $\frac{J_a}{\delta_{a1}}$, and be absorbed with amplitude
$(-A_1)$. This results in a next-NN interaction with strength
$\frac{-A_1}{\delta_{a1}} \frac{J_a}{\delta_{a1}} \frac{J_a}{%
  \delta_{a1}}(-A_1)=\frac{J_a^2}{%
  \delta _{a1}^3}A_1^2>0$. In Fig.2(a), similar processes occur and they can
be understood by replacing $A_1$ and $\delta_{a1}$ with $A_2$ and
$\delta_{a2}$ in the above discussion. It generates XX
interactions. In Fig.2(b), due to the site-dependent modulation of the
phases of the laser $\omega_2$ and $\omega_4$, the NN sites have
different amplitudes, $%
B_i$ and $(-B_i)$ $(i=1,2)$. So the effective couplings of the NN and
next-NN interactions are $\frac{B_i}{\delta_{bi}}
\frac{J_b}{\delta_{bi}} (-B_i)=-\frac{J_b}{\delta _{bi}^2}B_i^2<0$ and
$\frac{B_i%
}{\delta_{bi}} \frac{J_b}{\delta_{bi}} \frac{J_b}{\delta_{bi}}
B_i=\frac{J_b^2}{\delta _{bi}^3}B_i^2>0$, respectively.  All the
effective couplings have a factor of 2 because the atoms that emit and
absorb the virtual photon can be switched. We can clearly see that the
total interactions of the next-NN sites enhance while the total
interactions of the NN sites cancel, leading to the effective
Hamiltonian in Eq. ({\ref{eq:final_H}}).

It should be noticed that the effective Hamiltonian in Eq.
(\ref{eq:final_H}) is highly tunable, because it is determined by
6 free parameters (2 parameters on frequency detuning and 4
parameters on intensity of control laser). As long as the large
detuning conditions are satisfied, the interaction strengths in
effective Hamiltonian Eq. (\ref{eq:final_H}) can be adjusted at
will allowing to realize arbitrary ratios between $J_1/J_2$ and
$\lambda_1/\lambda_2$. With its large realizable parameter space,
it is then convenient to use this model to simulate a large class
of XXZ Heisenberg spin chain problems in which competing
interactions between NN and next-NN play an essential role.

As an example, we discuss how our system can be used to simulate
frustration phenomena in 1d condensed matter physics. By introducing
a new stark shift\cite{Hartmann}, the effective local magnetic field
can be set to 0. When we choose system parameters such that
$J_2=\lambda_2>0$ and $J_1>0$, $\lambda_1>0 $, the Hamiltonian $H$
is the well-known anisotropic Heisenberg model with competing
interactions (``frustration") originally studied by Haldane
\cite{Haldane}. Haldane predicted four phases for this system at
zero temperature: spin fluid phase, N\'eel phase, dimer phase, and
bound phase. When the ratio of $J_2/J_1$ is increased, the
interesting phenomenon of spontaneous dimerization takes place.

In coupled cavities systems, addressing of individual cavities is
available. Therefore, one can also simulate a frustrated spin chain
with disorder. Yusuf et. al. predicted the random singlet phase and
the large spin phase in random antiferromagnetic spin-$1/2$ chains
with NN and next-NN couplings. A strong next-NN coupling will drive
the system to a large spin phase \cite{Yusuf}.

\begin{figure}[tbp]
\centerline{\includegraphics[height=.4\textheight]{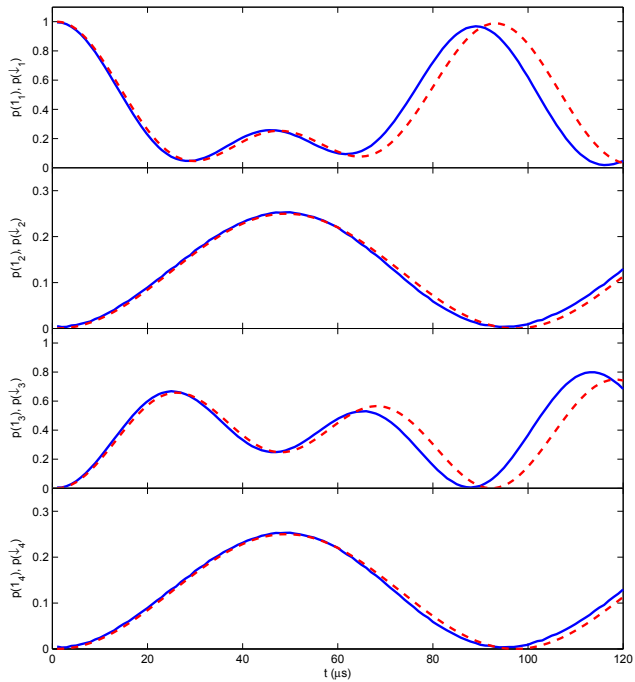}}
\caption{(color online). Time evolution of occupation probability
  $p(1_j)$ of state $%
  |1_j\rangle$  $(j=1,2,3,4)$, calculated using the full Hamiltonian (solid blue line) and
  effective spin Hamiltonian (dashed red line), for parameters $A_1=A_2=0.1$ GHz,
  $B_1=B_2=0.096$ GHz, $%
  A_3=B_3=0.02$ GHz, $\delta_1=4$ GHz, $\delta_2=3$ GHz, $\delta_3=1$
  GHz, $%
  g_a^2/\delta_{31}=g_b^2/\delta_{42}=0.1$ GHz, and $J_a=J_b=0.2$
  GHz.} \label{fig:simple}
\end{figure}

Aside from simulating frustrated spin problems of interests in
  the conventional context of condensed matter physics, our system
  with both tunable NN and next-NN interactions is also very valuable
  for quantum information studies. Spin chains with NN interactions
  are known to be useful for various quantum information tasks. For
  instance, they can be used to transfer quantum states \cite{Bose3}.
  Whether a spin chain with both NN and next-NN
  interactions can transfer quantum states more efficiently is an
  interesting topic to investigate, and our system provides a physical
  implementation for such a system readily. Perhaps more intriguingly,
  our system offers a powerful quantum simulation tool to study the
  deep connections between quantum entanglement and phase transitions
  in frustrated spin systems. This important subject has been
  discussed theoretically \cite{Nielsen, Lewenstein, Wang, Sun2},
  however no physical implementation of such a system with
  controllable NN and next-NN interaction strengths was available to
  check the findings in these preliminary theoretical research. Our
  system for the first time makes it possible to physically simulate
  relevant frustrated spin models. Because of the large realizable
  parameter space and ability to address individual spins in our
  system, we can simulate a large class of phase transitions and also
  study detailed information on entanglement in the system. This then
  allows to verify previous results on entanglement and phase
  transitions in frustrated spin systems and also simulate more
  complicated problems beyond the reach of conventional numerical
  methods.

Our derivation of the effective spin Hamiltonian in Eq.
(\ref{eq:final_H}) involves a large number of assumptions and
approximations. To check the validity of these approximations and the
accuracy of the results, we numerically simulate the dynamics
generated by the full Hamiltonian in Eq. (1) and the effective model
described by Eq. (\ref{eq:final_H}) and compare the results.

As an example, we consider four atoms in four cavities, initially
in the state $|1_1\rangle\otimes|2_2\rangle \otimes |2_3\rangle
\otimes |2_4\rangle$ corresponding to a spin state where only one
spin points down, $|\downarrow \uparrow \uparrow \uparrow
\rangle$. We calculate the time evolution of the probability of
atom $j$ in the $|1\rangle$ state, i.e., the occupation
probability $p(1_j)$ of state $ |1_j\rangle $ $(j=1,2,3,4)$ using
the full Hamiltonian in Eq. (1). This corresponds to the
probability of spin $j$ pointing down, $p(\downarrow_j)$, which we
calculate using the effective spin Hamiltonian in Eq.
(\ref{eq:final_H}). Fig.3 shows $p(1_j)$ and $p(\downarrow_j)$ for
an effective Hamiltonian(\ref{eq:final_H}) with
$J_1=0.0326$MHz, $J_2=0.0516$MHz, $%
\lambda_1=0.0187$MHz and $\lambda_2=0.0223$MHz \cite{localitem}.
To demonstrate the high tunability of our model, we have
deliberately chosen the system parameters such that next-NN
interactions are stronger than NN interactions. Due to periodic
boundary conditions, the results of $p(1_{2})$ and $p(1_{4})$ are
the same. As shown in Fig. 3, the numerical results of the
Hamiltonian Eq. (1) and those of the effective model agree with
each other reasonably well. Therefore, our effective model is
valid.

Quantum simulation based on our effective spin model is adversely
affected by nonidealities in the system, most notably the cavity
decay and limited lifetime of the excited atomic states. To
reliably observe exchange dynamics due to next-NN interactions,
the coupling strengths $J_{\sigma}$ and $\lambda_{\sigma}$ should
be at least one or two orders of magnitude larger than the cavity
decay rates and the lifetimes of the atomic upper levels. Defining
$\Omega=\max(\Omega_1, \Omega_2, \Omega_3, \Omega_4)$ , $g=\max
(g_a, g_b)$, $J=\max (J_a,J_b)$, $\Delta=\min(\Delta_{31},
\Delta_{32}, \Delta_{41}, \Delta_{42})$, and
$\delta=\min(\delta_1, \delta_2)$, one can see that the mean
population of the atomic excite state $|3 \rangle$ (or $|
4\rangle$) can be approximately written as $|\Omega/\Delta|^2$.
This results in an effective decay rate $|\Omega/\Delta|^2
\Gamma_E$ with $\Gamma_E$ the linewidth of the upper level.
Similarly, the effective cavity field decay rate can be expressed
as $|\Omega g/(\Delta \delta)|^2 \Gamma_C$, where $\Gamma_C$
describes the cavity decay of photons \cite{book}.  Since the
coupling $J_{\sigma}$ and $\lambda_{\sigma}$ is approximately
$4J^2 \Omega^2 g^2/(\Delta^2 \delta^3)$, this leads to the
constraints $\Gamma_E << 4 g^2 J^2/\delta^3$ and $
\Gamma_C<<4J^2/\delta$.  Without loss of generality, we assume
$\delta>2J$ with $2J$ the bandwidth of the photons in the cavity
array. Thus, to ensure that the photons tunnel between cavities
before they decay, the condition $\Gamma_C << J$ must be
satisfied. Additionally, for $\Gamma_E << 4 g^2 J^2/\delta^3$,
high cooperativity factors $g^2/(\Gamma_C \Gamma_E)$ and high
ratios of coupling to dissipation $g/\Gamma_E$ are favorable.

Experimentally, the requirements on parameters discussed above can be
fulfilled in microcavities with high quality factors. For toroidal
microcavities in \cite{Splillance}, the predicted critical atom number
can approach $10^{-7}$, which results in a cooperativity factor
$g^2/(\Gamma_C \Gamma_E) \sim 10^7$ and a ratio $g/\Gamma_E \sim
10^3$. For a strongly coupled single quantum dot-cavity system
\cite{Hennessy}, $g\simeq 20\mbox{GHz}$ can be achieved with
$\Gamma_C=24\mbox{GHz}$ and $\Gamma_E=18\mbox{MHz}$, for which we have
$g^2/(\Gamma_C \Gamma_E)\sim 10^3$ and $g/\Gamma_E \sim 10^3$. In both
systems, the photons can transfer between different cavities via their
evanescent fields or optical fibres. Thus, our effective spin model
with next-NN interactions can be readily and reliably realized in
presently available systems.

In this work, we have shown that one dimensional frustrated spin
models can be simulated using atoms in photon coupled cavities. By
choosing the phases of control lasers appropriately, it is possible to
make interactions arising from different level structures add up or
cancel each other, which then allows to adjust the ratio between NN
and next-NN interaction strengths at will. Quantum simulation using
our system is robust and reliable since the atoms are only virtually
excited and no real photon absorption and generation are involved and
thus dissipative processes such as photon decay and spontaneous
emission are strongly suppressed.

This work was funded by National Fundamental Research Program
(2006CB921900), the Innovation funds from Chinese Academy of
Sciences, and National Natural Science Foundation of China (Grant
No. 60621064, No. 10874170, No. 10875110, and No. 60836001). Z.-W.
Zhou gratefully acknowledges the support of K.C.Wong Education
Foundation, Hong Kong.


\begin{references}

\bibitem{Jaksch} D. Jaksch, C. Bruder, J. I. Cirac, C. W. Gardiner, and P.
Zoller, Phys. Rev. Lett. \textbf{81}, 3108 (1998).

\bibitem{Greiner} M. Greiner, O. Mandel, T. Esslinger, T. W. H\"onsch, and
I. Bloch, Nature (London) \textbf{415}, 39 (2002).

\bibitem{Calarco} T. Calarco, U. Dorner, P. Julienne, C. Williams, and P. Zoller,
Phys. Rev. A \textbf{70}, 012306 (2004); K. G. H. Vollbrecht, E.
Solano, and J. I. Cirac, Phys. Rev. Lett. \textbf{93}, 220502
(2004); J. Joo, Y. L. Lim, A. Beige, and P. L. Knight, Phys. Rev. A
\textbf{74}, 042344 (2006); C. Zhang, S. L. Rolston, and S. D.
Sarma, Phys. Rev. A \textbf{74}, 042316 (2006); Z. W. Zhou, Y. J.
Han, G. C. Guo, Phys. Rev. A \textbf{74}, 052334 (2006).

\bibitem{Plenio1} M. J. Hartmann, F. G. S. L. Brand\~ao, and M. B.
Plenio, Laser Photon. Rev. \textbf{2}, 527 (2008)

\bibitem{Plenio} M. J. Hartmann, F. G. S. L. Brand\~ao, and M. B. Plenio, Nat. Phys.
\textbf{2}, 849 (2006).

\bibitem{Greentree} A. D. Greentree, C. Tahan, J. H. Cole, and L. C. L. Hollenberg,
Nat. Phys. \textbf{2}, 856 (2006).

\bibitem{Angelakis} D. G. Angelakis, M. F. Santos, and S. Bose, Phys. Rev. A \textbf{76},
031805 (R) (2007).

\bibitem{Hartmann} M. J. Hartmann, F. G. S. L. Brand\~ao, and M. B. Plenio, Phys. Rev.
Lett. \textbf{99}, 160501 (2007).

\bibitem{Cho} J. Cho, D. G. Angelakis, and S. Bose, Phys. Rev. A \textbf{78}, 062338
(2008).

\bibitem{Liu} A. C. Ji, X. C. Xie, and W. M. Liu, Phys. Rev. Lett. \textbf{99}, 183602
(2007).

\bibitem{Kay} A. Kay and D. G. Angelakis, Europhys. Lett. \textbf{84}, 20001 (2008).

\bibitem{Fazio} D. Rossini and R. Fazio, Phys. Rev. Lett. \textbf{99}, 186401 (2007).

\bibitem{Bose2} J. Cho, D. G. Angelakis, and S. Bose, Phys. Rev. Lett. \textbf{101}, 246809
(2008).

\bibitem{Bose1} J. Cho, D. G. Angelakis, and S. Bose, Phys. Rev. A \textbf{78}, 022323
(2008).

\bibitem{Sun} Y. Li, M. X. Huo, Z. Song, and C. P. Sun, arXiv:0802.0079.


\bibitem{Zhou} L. Zhou, W. B. Yan, and X. Y. Zhao, J. Phys. B, At.
Mol. Opt. Phys. \textbf{42}, 065502 (2009)

\bibitem{Splillance} S.M. Splillance \emph{et al.}, Phys. Rev. A \textbf{71},013817 (2005).

\bibitem{Hennessy} K. Hennessy \emph{et al.}, Nature (London) \textbf{445}, 896 (2007); B.S.
  Song \emph{et al.}, Nat. Mater. \textbf{4}, 207 (2005).

\bibitem{Wallraff} A. Wallraff \emph{et al.}, Nature (London) \textbf{431},
 162 (2004)

\bibitem{Schollwock} U. Schollw\"ock \emph{et} \emph{al}., \emph{Quantum
Magnetism} (Springer-Verlag Berlin Heidelberg 2004).

\bibitem{Sachdev} S. Sachdev, \emph{Quantum Phase Transitions} (Cambridge University Press, Cambridge, England, 1999).

\bibitem{Haldane} F. D. M. Haldane, Phys. Rev. B \textbf{25}, 4925 (1982).

\bibitem{Somma} R. D. Somma and A. A. Aligia, Phys. Rev. B \textbf{64}, 024410 (2001).

\bibitem{Zhao} J. Zhao, D. X. Yao, S. Li, \emph{et} \emph{al}., Phys. Rev. Lett. \textbf{101}, 167203
(2008).

\bibitem{Vedral} L. Amico, R. Fazio, A. Osterloh and V. Vedral, Rev. Mod.
Phys. \textbf{80}, 518 (2008).

\bibitem{Nielsen} C. M. Dawson, and M. A. Nielsen1, Phys. Rev. A \textbf{69}, 052316
(2004).

\bibitem{Lewenstein} A. Sen(De), U. Sen, J. Dziarmaga, A. Sanpera, and M.
Lewenstein, Phys. Rev. Lett. \textbf{101}, 187202 (2008).

\bibitem{Wang} Y. Chen, Z. D. Wang, and F. C. Zhang, Phys. Rev. B \textbf{73}, 224414
(2006).

\bibitem{Sun2} X. F. Qian, T. Shi, Y. Li, Z. Song and C. P. Sun, Phys. Rev. A \textbf{72}, 012333 (2005).

\bibitem{James} D. F. V. James and J. Jerke, Can. J. Phys. \textbf{85}, 625 (2007).

\bibitem{mag} $h_j\approx (\frac{\Omega _1 ^2}{\Delta _{31}}+\frac{\Omega _4 ^2}{\Delta _{41}})-(\frac{\Omega _3 ^2}{\Delta _{32}}+\frac{\Omega _2 ^2}{\Delta _{42}})+\frac 2{\delta _3}(B_3^2-A_3^2)+\sum_{p=a,b}\{\varsigma _{p1} [\frac 1{\delta _{p1}}(1+\frac{4J_p^2}{\delta _{p1}^2})+\kappa _p\frac{2J_p}{\delta _{p1}^2}]+\frac{\varsigma _{p2}}{\delta _{p2}}(1+\frac{2J_p^2}{\delta _{p2}^2})\}$, here, $\varsigma _{a1}=-A_1^2,\varsigma _{b1}=B_1^2,\varsigma _{a2}=A_2^2,\varsigma _{b2}=-B_2^2,\kappa _a=1,\kappa _b=-1.$


\bibitem{Yusuf} E. Yusuf and K. Yang, Phys. Rev. B \textbf{68}, 024425
(2003).

\bibitem{Bose3} S. Bose, Contemp. Phys. \textbf{48}, 13 (2007); Z. Song
and C. P. Sun, Low Temp. Phys. \textbf{31}, 686 (2005).

\bibitem{localitem} Here, local effective magnetic field item is
omitted due to the commutation between the XXZ type of Heisenberg
interaction and the homogeneous magnetic field.

\bibitem{book} M.O. Scully and M.S. Zubairy, \emph{Quantum Optics} (Cambridge University Press, Cambridge, England, 1997).

\end{references}
\end{document}